\documentstyle[11pt,newpasp,twoside,epsf]{article}
\def\edcomment#1{\iffalse\marginpar{\raggedright\sl#1\/}\else\relax\fi}
\reversemarginpar

\begin{document}
\title{The Formation and Early Evolution of Protostellar
Accretion Disks}
\author{Arieh K\"onigl}
\affil{Department of Astronomy \& Astrophysics, University of
Chicago, 5640 S. Ellis Ave., Chicago, IL 60637, U.S.A.}
\begin{abstract}
Newly formed stars are often observed to possess circumstellar
disks, from which mass continues to be accreted onto the star
and fed into outflowing jets, and which eventually may evolve
into dusty debris disks and planetary systems. Recent modeling
developments have made it possible, for the first time, to study
the formation and early evolution of rotationally supported
protostellar disks in the context of a realistic scenario of
star formation in weakly ionized, magnetic, molecular cloud
cores. The derived semianalytic solutions incorporate
ambipolar diffusion and magnetic braking and may be
extended to include centrifugally driven disk
winds; they can be used to examine the full range of
expected behaviors of real systems and their dependence on
physical parameters.
\end{abstract}

\section{Introduction}
Protostellar disks are of much interest in the study of star
formation since it is likely that most
of the mass assembled in a typical low-mass young stellar object (YSO) is
accreted through a disk. Furthermore, such disks are the incubators of planetary
systems, so their properties are also directly relevant to the
process of planet formation. Circumstellar disks have been detected in
$\sim 25-50\%$ of pre--main-sequence stars in nearby dark
clouds (e.g., Beckwith \& Sargent 1993). At the time when the
protostars become visible as classical T Tauri stars, their disk
masses (as inferred from observations of
dust emission and from spectroscopic measurements) are usually
$\la 10\%$ of the central mass.

Rotationally supported circumstellar disks evidently originate
in the collapse of self-gravitating, rotating, molecular cloud
cores. Molecular line observations (e.g., Goodman et al. 1993;
Kane \& Clemens 1997) have established  that a majority of dense
($\ga 10^4\, {\rm cm}^{-3}$) cloud cores show evidence of
rotation, with angular velocities $\sim
3{\times}10^{-15}-10^{-13}\, {\rm s}^{-1}$ that tend
to be uniform on scales of $\sim 0.1\, {\rm pc}$, and
with specific angular momenta in the range $\sim
4{\times}10^{20}-3{\times}10^{22}\, {\rm cm}^2\, {\rm s}^{-1}$.
The cores can transfer angular momentum to the ambient gas
through torsional Alfv\'en waves (by the process of {\em magnetic
braking}), and this mechanism also acts to align their angular
momentum vectors with the local large-scale magnetic field
(e.g., Mouschovias \& Ciolek 1999). This alignment can occur on
the dynamical timescale and hence could be achieved even in
cores whose lifetimes are of that order (as in the
hierarchical-ISM scenario of Elmegreen 2000). Once dynamical
collapse is initiated and a core goes into a
near--free-fall state, the
specific angular momentum is expected to be approximately
conserved, resulting in a progressive increase in the
centrifugal force that eventually halts the collapse and gives
rise to a rotationally supported disk on scales $\sim 10^2\,
{\rm AU}$. These expectations are
consistent with the results of molecular-line interferometric
observations of contracting cloud cores (e.g., Ohashi et
al. 1997; Belloche et al. 2002).

This contribution summarizes recent advances in {\em
semianalytic} modeling of the formation and early evolution of
protostellar accretion disks. It has now become possible to
study these processes in the context of a realistic scenario of
star formation in molecular cloud cores that are threaded by a
dynamically significant magnetic field. A more detailed account
of this work is given in Krasnopolsky \& K\"onigl (2002;
hereafter KK02).

\section{Modeling Framework}

Numerical simulations of magnetically supported clouds have demonstrated that the
gas rapidly contracts along the field lines and maintains force
equilibrium along the field even during the collapse phase
(e.g., Fiedler \& Mouschovias 1993; Galli \& Shu 1993), including in
cases where the clouds are initially elongated in the field
direction (e.g., Nakamura, Hanawa, \& Nakano 1995; Tomisaka 1996). This
motivates treating the collapse as being quasi
one-dimensional.\footnote{The complementary approach of axisymmetric, 2D
collapse simulations in which new mass is added to the system
only from above and below the disk plane tends to produce
disk-to-YSO mass ratios $\sim 1$, much higher than typically observed.}

To obtain semianalytic solutions, KK02 adopted the assumption of
{\em self-similarity} in space and time, with a similarity
variable $x\equiv r/Ct$ (where $r$ is the distance from the
origin, $C$ is the isothermal speed of
sound, and $t$ is the time).\footnote{For a typical sound speed 
$C= 0.19\ {\rm km\ s^{-1}}$, $x=1 \Leftrightarrow \{400,\
4000\}\ {\rm AU}$ at $t=\{10^4,\  10^5\}\ {\rm yr}$.} This
assumption is motivated by the fact that core collapse is a
{\em multiscale} problem, which is expected to
assume a self-similar form away from the outer and inner
boundaries and not too close to the onset time (e.g., Shu 1977;
Hunter 1977). This behavior has been
verified  by previous numerical and semianalytic treatments of restricted
core-collapse problems -- with/without rotation and with/without
magnetic fields. The assumption of isothermality, which
underlies the $C= const$ ansatz, is justified mainly by
the fact that thermal stresses do not
play a major role in the dynamics of the collapsing core.

Molecular cloud cores are known to be weakly ionized. Therefore,
even though they also carried out reference calculations under
the assumption of ideal MHD, KK02 incorporated {\rm ambipolar
diffusion} into the model: the magnetic field lines are frozen
into the charged particle component (ions, electrons, grains)
and couple to the dominant neutral component through
ion--neutral drift. Although the drift velocity is
negligible during the early phase of the dynamical collapse,
ambipolar diffusion becomes important within
the gravitational ``sphere of influence'' of the YSO once the
central mass begins to grow (Ciolek \& K\"onigl 1998;
Contopoulos, Ciolek, \& K\"onigl 1998). When the incoming matter enters this region, it
decouples from the field and continues moving
inward. The decoupling front, in turn, moves outward and steepens into a
C-type {\em ambipolar diffusion} (AD) {\em shock} (the existence of which was first
predicted by Li \& McKee 1996).

To incorporate ambipolar diffusion
into the self-similarity formulation, it is necessary to assume
that the ion density scales as the square root of the neutral
density: $\rho_i = K \rho^{1/2}$. As discussed in KK02, this is
a good approximation for the core-collapse
problem: it applies on both ends of a density range spanning
$\sim 8$ orders of magnitude, which applies roughly on radial
scales $\sim 10-10^4\ {\rm AU}$, with $K$ varying by only 1 order of
magnitude across this interval.

The transition from a nearly freely
falling, collapsing core to a quasi-\break
stationary, rotationally supported disk involves a strong deceleration in a\break
{\em centrifugal shock}. This shock is  distinct
from the ambipolar-diffusion shock mentioned above: it typically
occurs at a different radius and is hydrodynamic,
rather than hydromagnetic, in nature.

To allow mass to accumulate at the center in a 1D, rotating-core collapse, an
angular momentum transport mechanism must be present. In their
basic model, KK02 assumed that {\em vertical} transport through
magnetic braking continues to operate also during the collapse phase of the core
evolution. To incorporate this mechanism into the self-similar
model, it is necessary to assume that $V_{\rm A,ext}$, the Alfv\'en speed in the
external medium, is a constant.\footnote{A nearly constant value $V_{\rm A,ext}
\approx 1\ {\rm km\ s^{-1}}$ is, in fact, indicated in
molecular clouds
in the density range $\sim 10^3-10^7\ {\rm cm}^{-3}$ (e.g., Crutcher 1999).}
KK02 verified that, in their derived solutions, magnetic braking indeed
dominates the most likely alternative angular-momentum transport
mechanisms --- magnetorotational
instability-induced turbulence and gravitational
torques. However, they also found that angular momentum
transport by a {\em centrifugally driven magnetic disk wind}
arises naturally (and may dominate) in their fiducial disk
solutions. They went on to show that the latter mechanism may be
incorporated into the model without significantly modifying the
basic formulation.

\section{Self-Similar Model}

KK02 employed a set of vertically-integrated thin-disk
equations. In their formulation (using cylindrical coordinates
$r,\, \phi,\, z$), they took the vertical magnetic field component to be
constant (except when evaluating $\partial B_z/\partial z$)
and assumed that $B_r$ and $B_\phi$ increase $\propto z$ in the disk; all terms
${\mathcal{O}}(H/r)$ (where $H$ is the disk scale height) were
neglected except in the azimuthal current-density term $[B_{r,s}-H({\partial
B_z}/{\partial r})]$ (where the subscript $s$ denotes a surface
value). Furthermore, a monopole approximation was employed for
the radial gravity $g_r$ and for $B_{r,s}$ (e.g., Li \& Shu 1997): 
$g_r = GM(r,t)/r^2$, $B_{r,s}=\Psi(r,t)/2\pi r^2$, relating
these quantities to, respectively, the mass and magnetic flux
enclosed within $r$.

In the self-similar formulation, the various physical quantities
are expressed as dimensionless functions of the similarity
variable $x$ in the following fashion:
\begin{displaymath}
H(r,t)=Ct\,h(x)
\, ,\ \ \Sigma(r,t)=(C/2\pi Gt)\,\sigma(x)\ ,
\end{displaymath}
\begin{displaymath}
V_r(r,t)=C\,u(x)
\, ,\ \ V_\phi(r,t)=C\,v(x)\ ,
\end{displaymath}
\begin{displaymath}
g_r(r,t)=(C/t)\,g(x)
\, ,\ \ J(r,t)=C^2t\,j(x)\ ,
\end{displaymath}
\begin{displaymath}
M(r,t)=(C^3t/G)\,m(x)\, ,
\ \ \dot{M}(r,t)=(C^3/G)\,\dot{m}(x)\ ,
\end{displaymath}
\begin{displaymath}
{\mathbf{B}}(r,t)=(C/G^{1/2}t)\,{\mathbf{b}}(x)\, ,
\Psi(r,t)=(2\pi C^3t/G^{1/2})\psi(x)\ ,
\end{displaymath}
where $\Sigma$ is the surface mass density, ${\mathbf{V}}$ is the
velocity, $J$ is the specific angular momentum, and $\dot M$ is
the mass accretion rate. From the assumption of vertical
hydrostatic equilibrium, one can derive a quadratic equation for
$h(x)$, whose solution is
\begin{displaymath}
h=\frac{\hat \sigma x^3}{2\hat m_c}
\left[-1 + \left(1+\frac{8\hat m_c}{x^3\hat
\sigma^2}\right)^{1/2}\right]\ ,
\end{displaymath}
where $\hat m_c\equiv m_c-x^3b_{r,s}(db_z/dx)/\sigma$ (with
$m_c$ being the central-mass eigenvalue) and $\hat
\sigma \equiv \sigma + (b_{r,s}^2+b_{\phi,s}^2)/\sigma$.

The adopted initial conditions were
$\sigma\rightarrow\frac{A}{x},\ b_z\rightarrow
{\sigma}/{\mu_0},\ u\rightarrow u_0,\ v\rightarrow v_0$ as
$x\rightarrow \infty$, 
with the parameter values $A=3$, $\mu_0=2.9$, $u_0=-1$  based on numerical
simulations of nonrotating magnetic cores and with the choice of the
value of $v_0$ motivated by the measured angular velocities in
molecular cloud cores.

The behavior in the limit $x\rightarrow 0$ (corresponding to
$r\rightarrow 0$ at a fixed $t$) can be derived from the
constituent equations. In particular, the asymptotic behavior of
an ambipolar diffusion-dominated circumstellar disk is given by
\begin{eqnarray}
\dot m&=&m=m_c\ ,\nonumber\\
j&=&m_c^{1/2}x^{1/2}\ ,\nonumber\\
-u&=&w=(m_c/\sigma_1)x^{1/2}\ ,\nonumber\\
\sigma&=&
\frac{(2\eta/3\delta)(2m_c)^{1/2}}{[1+(2\eta/3\delta)^{-2}]^{1/2}}
x^{-3/2}\nonumber\\
&\equiv& \sigma_1x^{-3/2}\ ,\nonumber\\
b_z&=&-b_{\phi,s}/\delta=
[{m_c}^{3/4}/(2\delta)^{1/2}]x^{-5/4}\ ,\nonumber\\
b_{r,s}&=&\psi/x^2=(4/3)b_z\ ,\nonumber\\
h&=&\{2/[1+(2\eta/3\delta)^2]m_c\}^{1/2}x^{3/2}\ .\nonumber
\end{eqnarray}

There are 4 model parameters that can be varied to explore the
solution space: $\delta\equiv |B_{\phi,s} |/B_z$ (the adopted
cap on the azimuthal field strength), $\alpha \equiv
C/V_{\rm A,ext}$, $v_0\equiv V_{\phi,0}/C$, and $\eta \equiv
\tau_{\rm ni} (4\pi G\rho)^{1/2}$ (where $\tau_{ni}\propto
1/\rho_i$ is the neutral--ion momentum-exchange time).
\section{Results}
\subsection{Fiducial Solution}
This solution corresponds to $\eta=1$, $v_0=0.73$, $\alpha =
0.08$, and $\delta=1$ (which yield $m_c = 4.7$). In this case
the initial rotation is not
very fast and the braking is moderate, leading to the formation
of a disk (with outer boundary at the centrifugal shock radius
$x_c=1.3\times 10^{-2}$) within the ambipolar-diffusion region (enclosed by the
AD shock radius $x_a=0.41 \approx
30\, x_c$). One can distinguish the following main flow regimes
(see Fig. 1):
\newpage
\begin{figure}
\begin{center}
\leavevmode
\epsfysize 12cm
\epsfbox{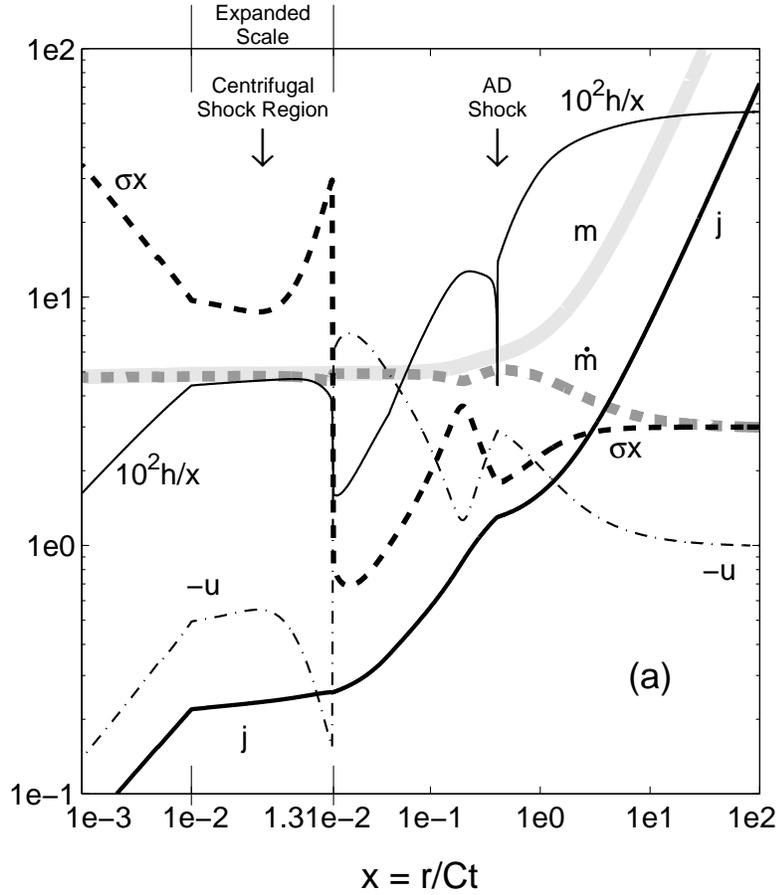}
\end{center}
\caption{Behavior of normalized flow variables in
the fiducial solution.}
\end{figure}
\begin{itemize}
\item 
Outer region ($x>x_a$): ideal-MHD infall.
\item
AD shock: resolved as a continuous transition (but
may in some cases contain a viscous subshock); KK02 estimated
$x_a\approx \sqrt{2}\eta/\mu_0.$
\item 
Ambipolar diffusion-dominated infall ($x_c < x <
x_a$): near free-fall controlled by the central YSO's gravity.
\item 
Centrifugal shock: its
location depends sensitively on the diffusivity parameter
$\eta$, which affects the amount of magnetic braking for
$x<x_a$; KK02 estimated $x_c\approx (m_cv_0^2/A^2)\exp
[-(2^{3/2}m_c/\mu_0)^{1/2}\eta^{-3/2}]$.
\item 
Keplerian disk ($x<x_c$): asymptotic behavior is approached after a
transition zone representing a massive ring (of width $\sim 0.1\, x_c$ and mass
$\sim 8\%$ of the disk mass within $x_c$, which in turn is $\la
5\%$ of $m_c$).
\end{itemize}
The asymptotic $x\rightarrow0$ solution (see \S~3) implies that the angle
between the meridional projection of ${\mathbf{B}}$ and the
rotation axis is radially constant and equal to  $\sim
53^{\circ}$, which exceeds the minimum value of $30^{\circ}$ for
launching a centrifugally driven wind from a ``cold'' Keplerian
disk (Blandford \& Payne 1982). This feature of the solution is
attractive in view of the fact that
centrifugally driven disk winds are a leading candidate for
the origin of the bipolar outflows that are frequently
observed to emanate from YSOs (e.g., K\"onigl \& Pudritz 2000).
\subsection{Limiting Cases: Fast Rotation and Strong Braking}
By modifying the model parameters, one can study the range of
possible behaviors in collapsing cores. Figure 2 shows two
limiting cases, which bracket the fiducial solution.
\begin{figure}[h]
\plottwo{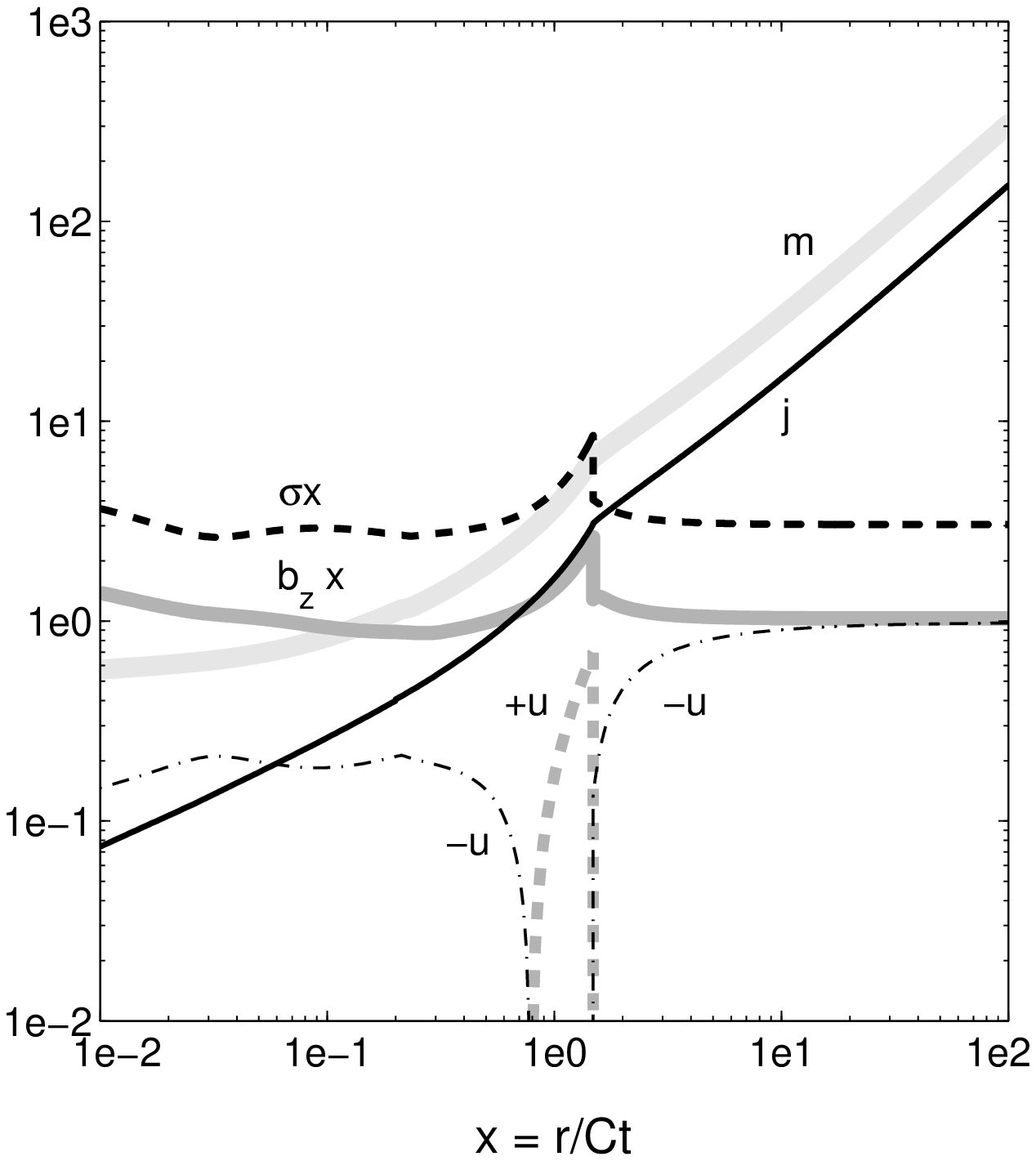}{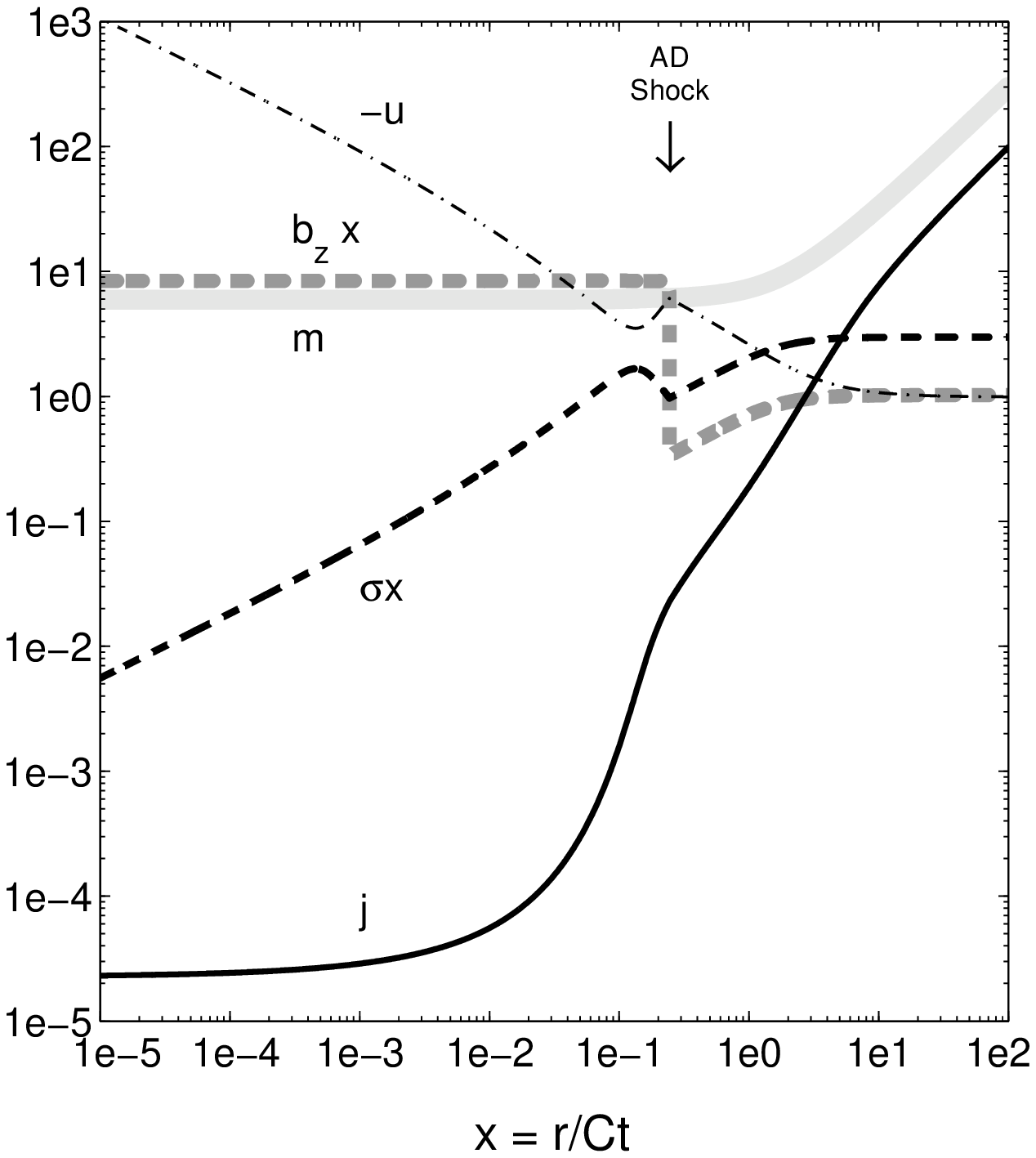}
\caption{Fast-rotation ({\bf left}) and strong-braking ({\bf
right}) solutions.}
\end{figure}

The {\em fast rotation} case differs from the fiducial solution
primarily in having a large initial-rotation parameter
($v_0=1.5$). It has the following distinguishing features:
\begin{itemize}
\item  The centrifugal shock is located within the
self-gravity--dominated (and ideal-MHD) region; a back-flowing
region is present just behind the shock.
\item The central mass is comparatively small ($m_c = 0.5$),
giving rise to a non-Keplerian outer disk region.
\item The ideal-MHD/ambipolar-diffusion transition
occurs behind the centrifugal shock and is gradual rather than sharp.
\end{itemize}
The {\em strong braking} case ($\eta=0.5$, $v_0=1$, $\alpha =
10$, and $\delta=10$; yielding $m_c=5.9$) is characterized by
large values of the braking parameters $\alpha$ and $\delta$. It
is distinguished by having
\begin{itemize}
\item no centrifugal shock (or circumstellar disk); the
$x\rightarrow\ 0$  behavior resembles that of the nonrotating
collapse solution of Contopoulos et al. (1998).
\end{itemize}
\section{Some Implications of the Model}
The formation of rotationally supported
circumstellar accretion disks basically resolves the angular momentum
problem in star formation (although the exact value of the YSO angular
momentum is determined by processes near the stellar surface
that are not included in this model). In particular, the derived
solutions demonstrate that angular momentum transport can be
sufficiently efficient to allow most of the inflowing mass to
end up (with effectively no angular momentum) at
the center, with the central mass dominating the dynamics well
beyond the outer edge of the disk even as the inflow is still in
progress. These solutions reveal that the ambipolar-diffusion shock,
even though it is usually located well outside the region where
the centrifugal force becomes important, helps to enhance
the efficiency of angular momentum transport through
the magnetic field amplification that it induces. The
revitalization of ambipolar diffusion behind the AD shock in turn
goes a long way toward resolving the magnetic flux problem
in star formation (as already pointed out by Ciolek \& K\"onigl 1998
and Contopoulos et al. 1998).

To the extent that self-similarity is a good approximation to
the situation in collapsing cloud cores, it is conceivable that
T Tauri (Class II) protostellar systems, whose disk masses are
typically inferred to be $\la 10\%$ of the central mass,
have had a similarly low disk-to-star mass ratio also during
their earlier (Class-0 and Class-I) evolutionary phases. It
would be interesting to test this possibility by observations. 
The model also predicts that, in cases where magnetic braking is
particularly strong, essentially all the angular momentum is
removed well before the inflowing gas reaches the center. Such
systems may correspond to slowly rotating YSOs that show no
evidence of a circumstellar disk (e.g., Stassun et al. 1999; 2001).
Another distinguishing characteristic of the solutions is the
appearance of an ambipolar-diffusion and/or a centrifugal
shock. The implied processing of the
disk material in these shocks (particularly the latter one)
may have implications to the composition of protoplanetary disks
(e.g., the annealing of silicate dust; see Harker \& Desch 2002).

The diffusive Keplerian disk models are by and large magnetorotationally
stable, basically because the matter/field
coupling is generally too weak to allow the instability to
grow. (The well-coupled surface layers should typically also be
stable because of strong magnetic squeezing; e.g., Wardle \&
K\"onigl 1993.) Furthermore, the Toomre stability
criterion to fragmentation ($Q_{\rm Toomre} > 1$) is well satisfied
for the rotationally supported disk solutions (except in the
outer layers of fast-rotation models). Angular momentum
transport by gravitational torques is unlikely to be important
under these conditions (e.g., Lin \& Pringle 1987). However, as noted
in \S~2 (see also \S~4.1), angular momentum transport by a
centrifugally driven wind may play a key role. KK02 found that
the steady-state, radially self-similar
disk-wind solution of Blandford \& Payne (1982) can be
naturally incorporated into the asymptotic ambipolar-diffusion
disk solution given in \S~3, making it possible to study the
effects of wind angular-momentum and mass removal from
the disk and to better constrain the relevant parameters of a
combined disk/wind model. In a preliminary analysis, they inferred that the
asymptotic solution evidently corresponds to the
weakly coupled disk/wind configurations discussed by Li (1996).
\acknowledgments
My work on star formation was initiated in 1980 when I came to Berkeley
as a postdoc and started interacting with Chris, Dave,
and Frank. Much of what I have learned about star
formation in the ISM during the subsequent years has
derived from our honorees' work and was inspired by their physical
insights and unique styles. I wish them all many years of
continued productive leadership in this field. The research reported in this
contribution was supported in part by NASA grant NAG5-3687.

\end{document}